\documentclass[runningheads]{llncs}
\usepackage{graphicx}
\usepackage{times}

\newcommand{\Xomit}[1]{ }
\newcommand{\pp}{\mathit{R}}

\newcommand{\cas}{\texttt{compare\&swap}}

\def\myview{\mathit{myview}}
\def\round{\mathit{round}}
\def\counter{\mathit{counter}}

\def\competitors{\mathit{competitors}}
\def\maximum{\mathit{max}}

\def\ffalse{\tt{false}}
\def\ttrue{\tt{true}}
\def\gcd{{\sf gcd}}
\def\return{{\sf return}}
\def\wait{{\sf wait}}
\def\done{{\tt done}}
\def\abort{{\tt abort}}
\def\acquire{{\sf acquire}}
\def\release{{\sf release}}
\def\mmax{{\sf max}}
\def\propose{{\sf propose}}
\def\mypref{\mathit{mypref}}

\begin{document}

\title{Fully Anonymous Shared Memory Algorithms}
%
\author{Michel Raynal\inst{1,2} \and Gadi Taubenfeld\inst{3}}
%
%
\institute{Univ Rennes IRISA, France \and
Department of Computing, Polytechnic University, Hong Kong
\and
The Interdisciplinary Center, Herzliya 46150, Israel
}
\maketitle              

\begin{abstract}
Process  anonymity has been studied for a long time.
Memory \linebreak anonymity is more recent. In an anonymous memory system,
there is no a priori agreement among the processes
on  the names of the shared registers they access.
As an example, a register named $A$
by a process $p$ and a shared register named $B$ by another
process $q$ may correspond to the very same register $X$,
while the same name $C$ may correspond to different register names
for the processes $p$ and $q$.
This article introduces the {\it fully anonymous} model, namely
a model in which both the processes and the registers are anonymous.
A fundamental question is then ``is this model meaningful?'',
which can be translated as ``can non-trivial  fundamental
problems be solved in such a very weak computing model?''

This paper answers this question positively. To this end, it
shows that mutual exclusion, consensus, and its weak version
called set agreement, can be solved despite full anonymity,
the first in a failure-free system, the others
in the presence of any number of process crashes.
More precisely, the paper presents three fully anonymous
algorithms. The first one is an $n$-process deadlock-free
mutual exclusion algorithm which assumes read/modify/write
registers. The model parameter $m$ defining the size of the
anonymous memory (number of registers), the paper also shows that
$m\in M(n)= \{~m~ \mbox{ such that }\forall~ \ell:~ 1 < \ell
\leq n$: $\gcd(\ell,m)=1\}$ is a necessary and sufficient
condition for the existence of such an algorithm.
Considering the same model in which any number of processes
may crash, an $n$-process wait-free  consensus algorithm is presented.
Finally, considering full anonymity and weaker registers
(namely, read/write registers)
an obstruction-free set agreement algorithm is presented.
As far as we know, this is the first time full anonymity is
considered, and where non-trivial concurrency-related problems are
solved in such a strong anonymity context.


\Xomit{
\Keywords:{Anonymity \and Anonymous shared memory \and
Anonymous processes \and Asynchrony \and Atomic read/write register \and Atomic
read/modify/write register \and Concurrency \and Consensus \and Crash failure \and
Mutual exclusion \and Set agreement \and
Synchronization \and Obstruction-freedom \and Wait-freedom.}
}
\end{abstract}

\section{Introduction:  Computing Model}

\subsection{On the process side}
\paragraph{Process anonymity}
The notion of {\it process anonymity} has been studied for a
long time from an algorithmic and computability point of view, both in
message-passing systems (e.g.,~\cite{A80,BR13,YK96}) and shared
memory systems (e.g.,~\cite{AGM02,BRS18,GR07}).
Process anonymity means that processes have no identity,
have the same code and the same initialization of their local variables
(otherwise they could be  distinguished).
Hence, in a process anonymous system, it is impossible
to distinguish a process from another process.

\paragraph{Process model}
The system is composed of a finite set of
$n\geq 2$ asynchronous,  anonymous sequential  processes denoted $p_1$, ..,
$p_n$.  Each process $p_i$ knows $n$, the number of processes, and $m$,
the number of registers.
The subscript $i$ in $p_i$ is only a notational convenience, which is not
known by the processes.
{\it Sequential} means that a process executes one step at a time.
{\it Asynchronous} means that each process proceeds in its own speed,
which may vary with time and always remains unknown to the other
processes.

\subsection{On the memory side}
\paragraph{Memory anonymity}
The notion of {\it memory anonymity} has been
recently introduced in~\cite{T17}.  Let us consider a
shared memory $R$ made up of $m$ atomic registers.  Such a
memory can be seen as an array with $m$ entries, namely $R[1..m]$.
In a non-anonymous memory system, for each index $x$, the name $R[x]$
denotes the same register whatever the process that accesses the
address $R[x]$. Hence in a non-anonymous memory,
there is an a priori agreement on the names of the shared
registers. This facilitates the implementation of
the coordination rules the processes have to follow to progress
without violating the safety properties associated with
the application they solve~\cite{HS08,R13,T06}.

The situation is different
in an anonymous memory, where there is no a priori agreement on the
name of each register.  Moreover, all the registers
of an anonymous memory are assumed to be
initialized to the same value (otherwise, their initial values could
provide information allowing processes to distinguish them). The
interested reader will find an introductory survey on process and
memory anonymity in~\cite{RC18}.

\paragraph{ Anonymous shared memory}
The shared memory is made up of $m\geq 1$ atomic anonymous
registers denoted $R[1...m]$. Hence, {\it all} the registers are
anonymous. As already indicated, due to its anonymity,
$R[x]$ does not necessarily indicate the same object for different processes.
More precisely,  a memory-anonymous system is such that:
\begin{itemize}
\item For each process $p_i$
  an adversary defined  a permutation $f_i()$  over the set
  $\{1,2,\cdots,m\}$, such that when $p_i$ uses the address
  $R[x]$, it actually accesses $R[f_i(x)]$,
\item No process knows the permutations, and
\item All the registers are initialized to the same
  default value denoted $\bot$.
\end{itemize}
%
\begin{table}[!ht]
\begin{center}
  \begin{tabular}{|c|c|c|}
    \hline
    identifiers  for an         &  local identifiers   & local identifiers    \\
    ~external observer~ & ~for process $p_i$~ &  ~for process $p_j$~ \\
    \hline
    \hline
    $R[1]$ &  $R_i[2]$    & $R_j[3]$  \\
    \hline
    $R[2]$  &  $R_i[3]$   & $R_j[1]$  \\
    \hline
    $R[3]$ &  $R_i[1]$  &  $R_j[2]$  \\
    \hline
    \hline
    permutation & $f_i():~[2,3,1]$ & $f_j():~[3,1,2]$\\
    \hline
  \end{tabular}
\end{center}
\caption{Illustration of an anonymous memory model}
\label{table:memory-anonymous}
\end{table}
%
An example of anonymous memory
is presented in Table~\ref{table:memory-anonymous}.
To make apparent the fact that $R[x]$ can have a different meaning for
different processes, we write  $R_i[x]$ when $p_i$ invokes $R[x]$.
\paragraph{Anonymous register model}
We consider two types of anonymous register models.
\begin{itemize}
\item RW (read/write) model.
  In this model all, the registers can be read or written by any process.
\item RMW (read/modify/write)  model.
In this model, each  register can be
read, written or accessed by an operation that atomically reads the
register and (according to the value read) possibly modifies it. More
precisely, this operation, denoted $\cas(R[x],old, new)$ has three
input parameters, a register $R[x]$ and two values $old$ and $new$,
and returns a Boolean value.  It has the following effect: if
$R[x]=old$ the value $new$ is assigned to $R[x]$ and the value
$\ttrue$ is returned (the $\cas()$ operation is then successful).
If $R[x]\neq old$, $R[x]$ is not modified, and
the value $\ffalse$ is returned.
\end{itemize}
In both models, {\it atomic}~\cite{L86} means that the operations on
the registers appear as if they have been executed sequentially, each
operation appearing between its start event and its end event, and for
any $x\in\{1,...m\}$, each read operation of a register $R[x]$ returns
the value $v$, where $v$ is the last value written in $R[x]$ by a
write or a successful $\cas(R[x],-,-)$ operation (we also say that the
execution is  {\it linearizable}~\cite{HW90}).
We notice that
the RMW model is at least as strong as the RW model.

On a practical side, it was recently shown that epigenetic cell
modifications can be modeled  by anonymous entities cooperating
through anonymous communication media~\cite{RTB18}.
Hence, fully anonymous distributed systems could inspire bio-informatics
(and be inspired by it)~\cite{NB2011,NB2015}.

\subsection{Content of the paper}
This article addresses mutual exclusion and agreement in fully anonymous
RMW and RW systems.
\paragraph{Mutual exclusion}
Mutual exclusion is the oldest and one of the most important
synchronization problems.  Formalized by E.W. Dijkstra in the
mid-sixties~\cite{D65}, it consists in building what is called a lock
(or mutex) object, defined by two operations, denoted $\acquire()$ and
$\release()$.
%
%
The invocation of these operations by a process $p_i$
follows the following pattern: ``$\acquire()$; {\it critical section};
$\release()$'', where ``critical section'' is any  sequence of code.
It is assumed that, once in the critical section,  a  process
eventually invokes $\release()$.
A mutex object must satisfy the following two properties.
\begin{itemize}
\item
Mutual exclusion:
No two processes are simultaneously in their critical section.
\item
Deadlock-freedom progress condition:  If there is a process
$p_i$ that has a pending operation $\acquire()$
(i.e., it invoked $\acquire()$ and its invocation is not terminated)
and there is no process in the critical section, there is a process $p_j$
(maybe $p_j\neq p_i$) that eventually enters the critical section.
\end{itemize}
Two memory-anonymous symmetric deadlock-free mutual exclusion algorithms
are presented in~\cite{ARW18}. One is for the RW register model,
the other one for the RMW register model.
These two algorithms are symmetric in the sense that the processes have identities
that can only be  compared for equality.
We notice that algorithms for anonymous processes are, by definition, symmetric.

Mutual exclusion cannot be solved in the presence of
process crash failures: if a process crashes just after it obtained the
critical section, it will never release it, and consequently the upper layer
application can block forever. The computing model must be enriched
with additional computability power (for example with failure detectors, see~e.g.,~\cite{BJ09,DFR19})
to be able to solve  mutual exclusion in the presence of failures.

\paragraph{Consensus}
Consensus is the most important agreement problem of fault-tolerant
distributed computing. Let us consider that any number of processes may
crash. A crash is a premature halting (hence, until it possibly crashes,
a process behaves correctly, i.e., reliably executes its code).
The consensus  problem consists in building  a one-shot operation,
denoted $\propose()$, which takes an input parameter (called {\it proposed}
value) and returns a result  (called {\it decided} value).
{\it One-shot} means that a process can invoke the operation at most once.
The meaning of this operation is defined as follows:
\begin{itemize}
\item
Validity: A decided value is a proposed value.
\item
Agreement: No two processes decide different values.
\item
Liveness (Wait-freedom): If a process does not crash, it decides a value.
\end{itemize}
Algorithms solving consensus in different types of non-anonymous shared memory
systems are described in several textbooks (e.g.,\cite{HS08,R13,T06}).
In this paper, we consider the  multi-valued version of consensus
(i.e., the domain of proposed values is not restricted to be binary).
While consensus can be solved from registers in a non-anonymous RMW
memory~\cite{H91}, it cannot in  a non-anonymous RW
memory~\cite{FLP85,LA87}.
It is, however, possible to solve
a weaker version of consensus  in non-anonymous RW system,
when the progress condition is weakened  as follows~\cite{HLM03}:
\begin{itemize}
\item
  Liveness (Obstruction-freedom): If a process does not crash,
  and executes alone during a long enough period, it decides.
  I.e., if a process runs alone starting from some
  point in the execution then it decides after executing a finite number of steps.
\end{itemize}

\paragraph{Set agreement}
Set agreement captures a weaker form of consensus in which the
agreement property is weakened as follows:
\begin{itemize}
\item At most $n-1$ different values are decided upon.
\end{itemize}
That is, in any given run, the size of the set of the decision values
is at most $n-1$.
In particular, in runs in which
the $n$ processes propose $n$ different values,
instead of forcing the processes to agree on a single value,
set agreement forces them to eliminate one of the proposed value.
The set agreement problem as defined above is also called
the $(n-1)$-set agreement problem~\cite{C93}.
While much weaker than consensus, as consensus, set agreement
cannot be solved in
non-anonymous RW memory systems~\cite{BG93,HS99,SZ00}
(and consequently cannot be solved in an anonymous memory either),
but, as consensus, it can be solved when
considering the weaker  obstruction-freedom progress condition.

\paragraph{Content of the paper}
Table~\ref{table:content} describes the technical content of the paper.
As an example, the first line associated with consensus
states that Section~\ref{sec:wf-consensus} presents a consensus algorithm
for an anonymous RMW system for $n>1$ and $m\geq1$.
As far as the mutex algorithm is concerned, it is also shown that
$m\in M(n)$, where $M(n)= \{~m~ \mbox{ such that }\forall~ \ell:~ 1 < \ell
\leq n$: $\gcd(\ell,m)=1\}$ is a \emph{necessary and sufficient} condition
on the size of the anonymous memory for such an algorithm.
\begin{table}[!ht]
\begin{center}
\begin{tabular}{|c|c|c|c|c|c|c|}
  \hline Problem & Section & Crashes &  Reg. type & Progress condition
                       & $n$ & $m$    \\
\hline
\hline
Mutual exclusion & \ref{sec:mutex} & No & RMW & Deadlock-freedom &  $n>1$ & $m\in M(n)$\\
\hline
Consensus & \ref{sec:wf-consensus} & Yes  & RMW & Wait-freedom
                                                 &  $n>1$  & $m\geq 1$\\
\hline
Set agreement &  \ref{sec:ob-set-agreement} & Yes
            & RW & Obstruction-freedom &  $n>1$  & $m\geq 3$\\
\hline
Consensus &  \ref{sec:ob-n=2-consensus}   & Yes & RW
                         & Obstruction-freedom &  $n=2$  & $m\geq 3$ \\
\hline
\end{tabular}
\end{center}
\caption{Structure of the paper}
\label{table:content}
\end{table}
%
\vspace{-1.0cm}
\section{Fully Anonymous Mutex using RMW Registers}
\label{sec:mutex}
As already mentioned, the mutual exclusion problem can be
solved for non-anonymous processes in both the anonymous RW register
model and the anonymous RMW register model~\cite{ARW18}.  However,
there is no mutual exclusion algorithm when the processes are
anonymous, even when using non-anonymous RW registers.  To see that,
simply consider an execution in which the anonymous processes run in
lock-steps (i.e., one after the other) and access the RW registers in
the same order.  In such a run it is not possible to break symmetry as
the local states of the processes will be exactly the same after each
such lock-step.

\subsection{A necessary and sufficient condition}
Let us recall that two integers $x$ and $y$ are said to be
\emph{relatively prime} if their greatest common divisor is 1,
notice that a number is \emph{not} relatively prime to itself.
Let $M(n)= \{ m \mbox{ such that }
           \forall \ell: 1< \ell \leq n: \gcd(\ell,m)=1\}$.

\begin{theorem}
\label{thm:RMW:mutex:manyProcesses}
There is a deadlock-free mutual exclusion algorithm for $n\geq 2$ anonymous
processes communicating through $m\geq 1$ anonymous {\em RMW} registers if
and only if $m\in M(n)$.
\end{theorem}

\proof
The proof of the  \emph{if direction},
follows from the very existence of the  deadlock-free
mutual exclusion algorithm for $n$ anonymous processes using $m$
anonymous RMW registers, where $m\in M(n)$, presented in
Section~\ref{sec:algo-mutex} and proved in Section~\ref{sec:proof-mutex}.
The proof of \emph{only if direction},
is an immediate consequence  of  the following observations:
\begin{itemize}
\item
The lower bound result
in~\cite{ARW18}, which  states that  $m\in M(n)$
is a necessary and sufficient condition for symmetric deadlock-free
mutual exclusion for $n$ non-anonymous processes
and  anonymous RMW registers.
As already noticed, algorithms for anonymous processes are, by definition, also symmetric.
\item
The non-anonymous processes and anonymous RMW registers model is at
least as strong as the fully anonymous RMW model.  \qed
\end{itemize}
\paragraph{Remark}
It is worth noticing that, from a
distributed computing understanding and computability point of view,
the condition $m\in M(n)$ shows that, as far as
deadlock-free mutual exclusion using RMW registers
is concerned, there is no computability gap between full anonymity
(as addressed here) and
register-restricted anonymity (addressed in~\cite{ARW18}).
 Both require $m\in M(n)$.  Actually
this condition tightly captures the initial ``asymmetry'' seed that allows
$n$ (anonymous or non-anonymous)  processes to solve deadlock-free mutex
using anonymous memory.

\subsection{A Fully anonymous RMW mutex algorithm}
\label{sec:algo-mutex}

\paragraph{The anonymous memory}
As already indicated,  each RMW register  of the anonymous memory $R[1..m]$,
is initialized to the value $\bot$. Moreover, it is assumed that $\bot$
is smaller than any non-negative integer.

\paragraph{Local variables at each process}
Each process $p_i$ manages the following local variables.
\begin{itemize}
\item $max_i$ is used to store the maximal value contained in a
  register (as seen by $p_i$).
\item $\counter_i$ is used to store the number of registers {\it owned}
  by $p_i$. A process {\it owns} a register when  it is the last
  process that wrote a non-$\bot$ value into this register.
\item $\myview_i[1..n]$ is an array of Boolean values,
  each initialized to $\ffalse$. When
  $\myview_i[j]$ is equal to $\ttrue$,  $p_i$ owns the register  $R_i[j]$.
\item $\round_i$ (initialized to $0$) is the round number
  (rung number in the ladder metaphor, see below) currently
  attained by $p_i$ in its competition to access the critical section.
  When $\round_i=n$, $p_i$ is the winner and can enter the critical section.
\end{itemize}

\paragraph{Principle of the algorithm:
           concurrent climbing of a narrowing  ladder}
At some abstract level, the principle that underlies the behavior of
the algorithm is simple.  Assume there is a ladder with $(n+1)$
rungs, numbered form $0$ to $n$.  Initially, all the processes are at
rung number $0$ (hence their local variables $\round_i$ are equal to
$0$).  For each process $p_i$, $\round_i$ is equal to the rung number
it attained.  The aim of the algorithm is to allow processes to
progress from a rung $r$ to the next rung $(r+1)$ of the ladder, while
ensuring that, for any $r\geq 1$, at most $(n-r+1)$ processes currently are at rung $r$.
From the local point of view of a process, this means that
process $p_i$ is allowed to progress to the rung $r=\round_i+1$ only when
some specific condition is satisfied.  This condition involves the
notion of {\it ownership} of an anonymous register (see above), and
the asymmetry assumption provided by the model, namely $m\in M(n)$.
\footnote{This principle is not new. As an example it is found
  in Peterson's $n$-process RW mutex algorithm, where processes raise
  and lower individual flags --visible by all processes-- and write
  their identity in a size $n$ non-anonymous memory \cite{P81}.}

\paragraph{Algorithm}
The algorithm is described in Fig.~\ref{fig:RMW:2mutex}.  A process
enters a ``repeat'' loop, that it will exit when it will have attained
the last rung of the ladder, i.e., when $round_i=n$.  When
$round_i=r>0$, which means $p_i$ is at round $r$, it attempts to own
more registers, by writing the rung number $r$ in the  registers
it owned previously and in new registers.  Its behavior in the loop body
is composed of three parts.
%
\begin{figure}[!p]
\small
\hrule
\vspace{0.1in}
\centering
\noindent
\textsc{Algorithm 1: code of an anonymous  process $p_i$}
%
\begin{tabbing}
\hspace{1.5em}\=\hspace{1.5em}\=\hspace{6em}\=\kill
\textbf{Constants:}\\
\> $n$, $m$: positive integers,  \`// \texttt{\# of processes}
\texttt{and \# of shared registers}\\
\> model constraint \`//
\texttt{$\forall~\ell:~1<\ell\leq n$, $m$ and $\ell$ are relatively prime}\\
\textbf{Anonymous RMW shared registers:}\\
\> $R[1..m]$: array of $m$ anonymous RMW registers, initially all $\bot$ \` // $\bot< 0$\\
\textbf{Local variables:}\\
\> $\myview_i[1..m]$: array of $m$ Boolean bits, initially all $\ffalse$ \`// \texttt{indicates ownership}\\
\> $\counter_i, \round_i, \maximum_i$: integer\\~\\
%
\hspace{1.5em}\=\hspace{.5em}\=\hspace{1.5em}\=\hspace{1.5em}\=\hspace{1.5em}\=\hspace{1.5em}\=\kill
{\bf operation} $\acquire()$ {\bf is} \\
1 \> $\counter_i\leftarrow 0; \round_i\leftarrow 0$\`// \texttt{begin entry code}\\
2 \> \textbf{repeat}\\
3 \>\> $\maximum_i\leftarrow 0$ \`// \texttt{check if another process is in a higher round}\\
4 \>\>
$\maximum_i \leftarrow  \mmax(max_i,R_i[1], \ldots, R_i[m])$ \`// \texttt{find maximum in $R_i[1..m]$}\\

5 \>\> \textbf{if} $\round_i < \maximum_i$ \textbf{then} $\round_i\leftarrow 0$ \` // \texttt{withdraw from the competition}\\
6 \>\> \textbf{else} $\round_i\leftarrow \round_i+1$ \textbf{fi} \`// \texttt{continue to the next round}\\
\\
7 \>\> \textbf{if} $\round_i =1$ \textbf{then} \`// \texttt{first round}\\
8 \>\>\> \textbf{for each} $j\in \{1,...,m\}$ \textbf{do} \`// \texttt{try to own as many shared}\\
9 \>\>\>\> $\myview_i[j]\leftarrow \cas(R_i[j],\bot,1)$  \`// \texttt{registers as possible}\\
10\>\>\>\> \textbf{if} $\myview_i[j]$ \textbf{then} $\counter_i\leftarrow \counter_i+1$ \textbf{fi} \textbf{od} \textbf{fi}\`// \texttt{own one more}\\
\\
11\>\> \textbf{if} $\round_i\geq 2$ \textbf{then} \`// \texttt{try to own additional released registers}\\
12\>\>\>  \textbf{for each} $j\in \{1,...,m\}$ \textbf{do} \\
13\>\>\>\> \textbf{if} $\myview_i[j]$ \textbf{then} $R_i[j]\leftarrow \round_i$  \textbf{fi} \textbf{od}\`// \texttt{update all owned registers}\\
14\>\>\> \textbf{for each} $j\in \{1,...,m\}$  \textbf{do} \\
15\>\>\>\> \textbf{while} $R_i[j]< \round_i$  \textbf{do} \`// \texttt{$R_i[j]< \round_i$ implies $\myview[j]=\ffalse$}\\
16\>\>\>\>\> $\myview_i[j]\leftarrow \cas(R_i[j],\bot,\round_i)$  \`// \texttt{try to own $R_ij]$}\\
17\>\>\>\>\> \textbf{if} $\myview_i[j]$ \textbf{then} $\counter_i\leftarrow \counter_i+1$ \textbf{fi} \textbf{od} \textbf{fi} \`// \texttt{own one more}\\
\\
18\>\> \textbf{if} $\round_i\geq 1$ \textbf{then} \`// \texttt{not eliminated}\\

19\>\>\> $\competitors \leftarrow n-\round_i +1$ \`// \texttt{max \# of competing processes}\\ 
20\>\>\> \textbf{if} $\counter_i < m / \competitors$ \textbf{then} \`// \texttt{withdraw from the competition}\\
21\>\>\>\>  \textbf{for each} $j\in \{1,...,m\}$ \textbf{do} \`// \texttt{since not own enough registers}\\
22\>\>\>\>\> \textbf{if} $\myview_i[j]$ \textbf{then} $R_i[j]\leftarrow \bot$; $\myview_i[j]\leftarrow \ffalse$  \textbf{fi} \textbf{od} \`// \texttt{release}\\ 

23\>\>\>\>   $\wait(\forall~j\in \{1,...,m\}:~R_i[j]=\bot)$;
\`//\texttt{wait until all are  $= \bot$}\\

24\>\>\>\> $\counter_i\leftarrow 0; \round_i\leftarrow 0$   \textbf{fi} \`// \texttt{start over}\\
25\> \textbf{until} $\round_i =n$ \`// \texttt{until the winner owns all $m$ registers}\\
26\> $\return(\done)$.\\~\\


{\bf operation} $\release()$ {\bf is } \\

27\> \textbf{for each} $j\in \{1,...,m\}$\textbf{do}
$R_i[j]\leftarrow \bot$; $\myview_i[j]\leftarrow \ffalse$ \textbf{od} \`// \texttt{release all}\\ 
28\> $\return(\done)$.

\end{tabbing}
\caption{\small{Deadlock-free mutual exclusion  for $n$
    anonymous processes and  $m\in M(n)$ anonymous RMW registers
\label{fig:RMW:2mutex}}}
\vspace{0.1in}
\hrule
\normalsize
\end{figure}
%
\begin{itemize}
\item
Part  1: lines~2-6.
A process $p_i$ first scans (asynchronously) all the registers to
know the highest value they contain. This value is stored in $max_i$
(lines~3-4). Then, if registers are  different from $\bot$
(i.e., are
owned by some processes, we have then $round_i< max_i$, line~5),
$p_i$ loops at lines~3-6 until
it finds all the registers equal to $\bot$.
In short, as $p_i$ sees that other processes climbed already at higher rungs,
it stays looping at the rung numbered $0$.
\item Part  2: lines~7-17. This part subdivides in two sub-parts,
  according to the round number of $p_i$.
  In both cases, $p_i$ tries to own as many registers as possible.
\begin{itemize}
\item $round_i=1$.  In this case, $p_i$ owns no registers. So, it
  scans the anonymous memory and, for each register $R_i[j]$, it
  invokes $\cas(R_i[j],\bot,1)$ to try to own it. If it succeeds,
  it updates $\myview_i$ and $counter_i$ (line~8-10).
\item $round_i\geq 2$. In this case, $p_i$ became the owner of some
  registers during previous rounds. It then confirms its ownership of
  these registers with respect to its progress to the current round
  $r$ (line~12-13).  Then it attempts to own more registers. But, to
  ensure deadlock-freedom, it considers only the registers that
  contain a round number  smaller than its current round $r$.
  The array $\myview_i$ and the local variable $counter_i$ are
  also updated according to the newly owned registers (line~14-17).
\end{itemize}
\item Part 3: lines~18-24.  The aim of this part is to ensure
  deadlock-freedom.  As the proof will show, if $p_i$ attains
  rung $r>0$ (i.e., $\round_i=r$), there are at most $(n-r+1)$ processes
  competing with $p_i$ (line~19), and these processes attained a rung
  $\geq r$. In this case, at least one of them (but not all) must
  withdraw from the competition so that at most $(n-r)$ processes
  compete for the rung $r$.

  The corresponding ``withdrawal''
  predicate is $\counter_i<m/(n-r+1)$ (line~20), which involves the
  asymmetry-related pair ($n,m)$ and $\round_i=r$, which  measures the
  current progress of $p_i$.  If the withdrawal predicate is false and
  $p_i$ attained $\round_i=n$,  it enters the critical section (predicate of
  line~25).  If the predicate is false and $\round_i<n$,   $p_i$
  re-enters the loop, to try to own more registers and progress to the
  next rung of the  ladder.

  If the withdrawal predicate is true, $p_i$ releases all the
  registers it owns and updates $\myview_i$ accordingly
  (lines~21-22). Then, it waits until it sees all the
  registers equal to their initial value (lines~23). After that, $p_i$
  resets its local variables to their initial values (lines~24),
  and re-enters the loop body.
\end{itemize}
\paragraph{Abortable mutex}
Let a {\it deadlock-free abortable} mutex algorithm be a mutex
algorithm that, while it always satisfies the deadlock-freedom
property, allows an invocation of $\acquire()$ to return the
control value $\abort$ in the presence of concurrency (see, e.g.,
~\cite{R13,T06}).  In this case, the invoking process $p_i$ learns
that the critical section is currently used by another process. From
its point of view, it is as if it did not invoke $\acquire()$.  Let us
observe that, with abortable mutex, all the invocations of
$\acquire()$ terminate (some obtaining $\done$ and others obtaining $\abort$).
The previous algorithm can be easily transformed
into a  deadlock-free  {\it abortable} algorithm by replacing
the statement $\round_i\leftarrow 0$  by $\return(\bot)$ at line~5, and
replacing the lines~23-24 by $\return(\abort)$.
\paragraph{Remark}
The algorithm remains correct if the predicate of line~25 is replaced with
the predicate  ``$\counter_i = m$'', namely once a process owns the
$m$ registers it may enter its critical section.
This shows that the algorithm establishes a strong
termination-related relation between the number of asynchronous rounds
(i.e., time)  and the size of the memory (i.e., space).
\subsection{Proof of the algorithm}
\label{sec:proof-mutex}
Reminder: $M(n)= \{ m \mbox{ such that } \forall \ell: 1< \ell \leq n:
\gcd(\ell,m)=1\}$.  Moreover, let us say that ``process $p_i$ executes
round $r$'' when its local variable $round_i=r$.
\begin{lemma}
\label{lemma-not-integer}
Let $m\in M(n)$ and  $r\in  \{2,..., n\}$.
The values  $m/(n-r+1)$ are  not integers.
\end{lemma}
\proof
The set of the values $(n-r+1)$ for $r\in \{1,...,n-1\}$
is  $X=\{n,n-1, ...,2\}$.
The fact that, for any  $x\in X$, $m/x$ is not an integer
is a direct consequence of the definition of $m$, namely,
$m\in M(n)$. \qed
%
\begin{lemma}
\label{lemma-invariant}
Let us consider the largest round $r$ executed
by processes. At most $(n-r+1)$ processes are  executing a round $r$.
\end{lemma}
\proof
Let us consider a process that executes line~6, where it sets
its local variable $\round_i$ to $1$. As there are $n$ processes,
trivially at most $n$ processes  are  simultaneously executing round $r=1$.
Let us assume (induction hypothesis) that round $r$ is the largest
round currently executed by processes, and at most $(n-r+1)$ processes
execute it.  We show that at most $(n-r)$ processes will execute round
$r+1$.

Let $P_r$ be the set of processes that execute round $r$.  Let
us consider the worst case, namely, $|P_r|=n-r+1$.  We have to show
that at least one process of $P_r$ will not execute round $(r+1)$.
This amounts to showing that at least one process $p_i$ of $P_r$ never exits
the wait statement of line~23, or executes line~24 where it resets its
variable $\round_i$ to $0$. Whatever the case, this amounts to showing
that there is at least one process $p_i$  of $P_r$ for which the predicate
$\counter_i<m/(n-r+1)$ is satisfied at line~20.

When a process of $P_r$ exits the set of statements of lines~8-10 when
$r=1$, or line~12-17 when $r>1$, the value of each anonymous register
is $\geq r$.  Let us observe that, when different from $0$, the local
variable $\counter_i$ of a process $p_i$ counts the number of
anonymous registers that this process set equal to $round_i$, where
$\round_i=r$, i.e.,
$\counter_i=|\{x\mbox{ such that } \myview_i[x]=\ttrue\}|$
(line~8-10 when $\round_i=1$, and lines~12-17 when $\round_i>1$).
Notice also that, in the last case, $\counter_i$ increases from round
to round and thanks to the atomicity of the operation
$\cas(\pp[j],\bot,\round_i)$ at line~9 or~16 that, with respect to the
registration in the local variables $\myview_i[1..n]$, no anonymous
register can be counted several times by the same process  or
counted by several processes.

Assume (by contradiction) that the predicate of line~20 is false
at each process of $P_r$, and let $\counter(x)$, for $1\leq x\leq
|P_r|$, be the value of their counter variables. Then
$\counter(1)+\cdots+\counter(|P_r|)=m$, and each counter
is greater or equal to $m/(n-r+1)$. Hence, $\forall
x:~\counter(x)\geq m/(n-r+1)$. As, due to
Lemma~\ref{lemma-not-integer}, $m/(n-r+1)$ is not an integer, it
follows that $\forall x:~\counter(x) \geq \lceil m/(n-r+1)\rceil$.
And consequently, $\counter(1)+\cdots+\counter(|P_r|) \geq (n-r+1)
\lceil m/(n-r+1)\rceil$.  But $(n-r+1) \lceil m/(n-r+1)\rceil>m$, a
contradiction.

Hence, at least one local variable $counter$ is such
that $counter < (m/(n-r+1)$. It follows that at least one process of
$P$ executes line~27, which concludes the proof of the lemma. \qed

\begin{theorem}
\label{theorem-mutex-safety}
No two processes are simultaneously in the critical section.
\end{theorem}
\proof
The theorem follows directly from
the previous lemma and the fact that a  process enters the critical
section only when its local variable $round =n$ (line~25). \qed

\begin{lemma}
\label{lemma-invariant-2}
Let $r$, $1 \leq r <n$, be  the highest round attained by processes.
At least one process attain the round $(r+1)$.
\end{lemma}
\proof
Let $r$, $1 \leq r <n$, be  the highest round attained by processes,
and $P(r)$ the corresponding set of processes.
As in the proof of Lemma~\ref{lemma-invariant},  let
$P_r$ be the set of processes that execute round  $r$.
As previously, we have  $\counter(1)+\cdots+\counter(|P_r|)=m$.
If the predicate of line~20 is satisfied at each process of $P_r$
we have $\forall x:~\counter(x) < m/(n-r+1)$. As due to
Lemma~\ref{lemma-not-integer}  $m/(n-r+1)$ is not an integer,
it follows that $\forall x:~\counter(x) \leq \lfloor m/(n-r+1)\rfloor $.
Consequently, $\counter(1)+\cdots+\counter(|P_r|)
\leq  (n-r+1)  \lfloor m/(n-r+1)\rfloor$.
But  $(n-r+1) \lfloor m/(n-r+1)\rfloor <m$, a contradiction.
\qed

\begin{theorem}
\label{theorem-mutex-liveness}
If at some time no process is inside the critical section and
one or more  processes want to enter the critical section,
at least one process will enter it.
\end{theorem}
\proof
The theorem follows directly from the previous lemma, applied from
round $1$ until round $n$. \qed

\section{Fully Anonymous Wait-free Consensus using RMW Registers}
\label{sec:wf-consensus}

When considering a fully anonymous system of size $m=1$, consensus can
be easily solved with the $\cas()$ operation: the first process that
writes its value in the single register $R[1]$ (initialized to $\bot$)
imposes it as the
decided value (actually, when $m=1$ the memory is not really anonymous).
%
When using anonymous objects, the fact that a given problem can be solved using
only one object (i.e., $m=1$) does \emph{not} imply that the problem can also be solved using
any finite number of $m\geq 1$ objects \cite{ARW18}.

The algorithm describes in Fig.~\ref{algo:consensus} presents a  simple
consensus algorithm for any size $m\geq 1$ of the
anonymous RMW memory.  This algorithm assumes that the set of
values that can be proposed is totally ordered.  Each process tries
to write the value it proposes into each anonymous register. Assuming
that at least one process that does not crash invokes $\propose()$,
there is a finite time after which, whatever the concurrency/failure
pattern, each anonymous register contains a proposed value. Then,
using the same deterministic rule  the processes decide the same value
(let us notice that there is an a priori statically defined agreement
on the deterministic rule used to select the decided value).
%
\begin{figure}[!t]
\small
\hrule
\vspace{0.1in}
\centering
\noindent
\textsc{Algorithm 2: code of an anonymous  process $p_i$}
\begin{tabbing}
\hspace{1.5em}\=\hspace{1.5em}\=\hspace{6em}\=\kill
\textbf{Constants:}\\
\> $n,m$: positive integers \`// \texttt{\# of processes}
\texttt{and \# of shared registers}\\
\textbf{Anonymous RMW registers:}\\
\> $R[1..m]$: array of $m$ RMW registers, initially  all $\bot$
 \`// \texttt{$\bot$ cannot be proposed}\\~\\
%
\hspace{1.5em}\=\hspace{.5em}\=\hspace{1.5em}\=\hspace{1.5em}\=\hspace{1.5em}\=\hspace{1.5em}\=\kill
       {\bf operation} $\propose(in_i)$ {\bf is}
                      \`// \texttt{$in_i$ value proposed by $p_i$}\\
1 \> \textbf{for each} $j\in \{1,...,m\}$  \textbf{do}
$\cas(R_i[j],\bot,in_i)$ \textbf{od} \`// \texttt{try to write}\\
2 \> $\return(\mmax(R_i[1],...,R_i [m]))$ \`//
\texttt{decide the max value in $R[1..m]$}.\\
\end{tabbing}
\vspace{-0.35cm}
\caption{\small{Consensus  for $n\geq 2$
    anonymous processes and  $m\geq 1$ anonymous RMW registers
    \label{algo:consensus}}}
\vspace{0.1in}
\hrule
\normalsize
\end{figure}
\section{Fully Anonymous Obstruction-free Set Agreement using RW Registers}
\label{sec:ob-set-agreement}
We present an obstruction-free  set agreement algorithm for
crash-prone anonymous $n$-process system, where communication is through
$m\geq 3$ anonymous RW registers.

\subsection{A fully anonymous RW set agreement algorithm}
The algorithm is described in Fig.~\ref{fig:ob-set-agreement}.
The anonymous memory is made up of  $m\geq 3$ RW atomic registers.
\begin{figure*}[t]
  \small
 \hrule
\vspace{0.1in}
\centering
\noindent
\textsc{Algorithm 3: code of an anonymous process $p_i$}
%
\begin{tabbing}
  \hspace{1.5em}\=\hspace{1.5em}\=\hspace{6em}\=\kill
\textbf{Constants:}\\
\> $n,m$: positive integers \`// \texttt{\# of processes} 
\texttt{and \# of shared registers}\\
\textbf{Anonymous RW registers:}\\
\> $R[1..m]$: array of $m$ anonymous RW registers, initially  all $\bot$
 \`// \texttt{$\bot$ cannot be proposed}\\
%
\textbf{Local variables:}\\
\> $\myview_i[1..m]$: array of $m$ variables\\
\> $\mypref_i$: integer; $j$: ranges over $\{0,...,m\}$
\end{tabbing}
\begin{tabbing}
  \hspace{1.5em}\=\hspace{1.5em}\=\hspace{1.5em}\=\hspace{1.5em}\=\kill

  {\bf operation} $\propose(in_i)$ {\bf is}
                      \`// \texttt{$in_i$ value proposed by $p_i$}\\

1 \> $\mypref_i \leftarrow in_i$\\
2 \> \textbf{repeat} \\ 
3 \>\> \textbf{repeat} \\ 
4 \>\>\> \textbf{for} $j=1$ \textbf{to} $m$ \textbf{do} $\myview_i[j]\leftarrow \pp_i[j]$ \textbf{od} \`\texttt{//read the shared array}\\
5 \>\>\> \textbf{if} $\exists\ value\neq \bot$ which appears in more than half of the entries of $\myview_i[1..m]$\\
6 \>\>\>\> \textbf{then} $\mypref_i \leftarrow value$ \textbf{fi}  \`\texttt{//update preference}\\
7 \>\>\>  $j \leftarrow$ \= an arbitrary index $k\in\{1,...,m\}$ such that $\myview_i[k]\neq\mypref_i$ \`// \texttt{search}\\
  \>\>\>\> or 0 if no such index exists \\
8 \>\>\>  \textbf{if} $j\neq 0$ \textbf{then} $\pp_i[j]\leftarrow \mypref_i$ \textbf{fi} \`// \texttt{write}\\
9 \>\> \textbf{until} $\forall j\in\{1,...,m\}: \myview_i[j]= \mypref_i$ \`// \texttt{my $\mypref_i$ is everywhere} \\
10 \>\> \textbf{for} $j=1$ \textbf{to} $m$ \textbf{do} $\myview_i[j]\leftarrow \pp_i[j]$ \textbf{od} \`// \texttt{read the shared array again}\\
11 \> \textbf{until} $\forall j\in\{1,...,m\}: \myview_i[j]= \mypref_i$ \`// \texttt{my $\mypref_i$ is everywhere} \\
12 \> $\return(\mypref_i)$. \`// \texttt{decide}
\end{tabbing}
\vspace{-0.3cm}
\caption{Fully anonymous obstruction-free set agr. algorithm for
  $n\geq 2$ proc. and   $m\geq 3$ registers}
\label{fig:ob-set-agreement}
\vspace{0.1in}
\hrule
\end{figure*}
%
Each anonymous RW register can store the preference of a process.
Each participating process $p_i$ scans the $m$ registers trying to write
its preference ($\mathit{preference}_i$) into each one of the $m$ registers.
Before each write, the process scans the shared array
(line 4), and operates as follows:
\begin{itemize}
\item
If its preference appears in all the $m$ registers (line~9), it reads
the array again (line~10), and if, for the second time, its preference
appears in all the $m$ registers (line~11), it decides on its
preference. 
\item
Otherwise, if some preference appears in more than half of the
registers (line~5), the process adopts this preference as its new
preference (line~6).
\end{itemize}
Afterward, the process finds some arbitrary entry in the shared array
that does not contain its preference (line~7) and writes it into that
entry (line~8).
Once the process finishes writing it repeats the above steps.
%
\subsection{Proof of the algorithm}
\begin{lemma}[Set agreement and  Termination under Obstruction-freedom]
 \label{lemma-sa-1}  
Any participating process that runs alone for a sufficiently long
time, eventually decides. Moreover, the processes that
decide, decide on at most $n-1$ different values.
\end{lemma}
\proof
Clearly in all the runs in which less than $n$ processes decide,
they decide on at most $n-1$ different values.
Below, we prove that in runs in which
all the $n$ processes participate and decide,
the $n$ processes decide on at most $n-1$ different values.

Let $\rho$ be an arbitrary run in which all the $n$ processes
participate and decide.  We prove that, in $\rho$, the processes
decide on at most $n-1$ different values.
Each one of the $n$ processes, \emph{before} deciding (line 12), must
first read all the $m$ registers (line 4), find out that its
preference appears in all the $m$ registers (line 9), then it must
read the array again (line 10) and only if, for the second time, its
preference appears in all the $m$ registers (line 11), it may decide
on its preference and terminate.  We call these last two consecutive
reads of the $m$ registers by a specific process a {\it successful
  double collect} (SDC) of that process.  We emphasize that from the moment
a process starts its successful double collect until it decides, the
process does not write.

Let us denote by $p_i$ and $p_j$ the \emph{last} two processes which start
their SDC in the run $\rho$. Clearly, by
definition, during these two last SDCs,
each one of the other processes has either decided and terminated or
has already started it SDC, and hence does not write during $p_i$ and
$p_j$ SDCs.
We  show that $p_i$ and $p_j$ must decide on the same value which
implies, as required, that the $n$ processes decide on at most $n-1$
different values in $\rho$.

From now on we focus only on processes $p_i$ and $p_j$.  Denote the
value that $p_i$ decides on by $v$.  Assume w.l.o.g.\ that $p_i$ has
started its (last and only) SDC before process $p_j$ has started its
(last and only) SDC.
Let $t_0, t_1$ and $t_2$ denote: the last time $p_i$ enters the inner
loop just before reading the $m$ registers (between lines 3-4), the
last time at which $p$ exits the inner loop (between lines 9-10), and
the time at which $p_i$ exits the outer loop (between lines 11-12),
respectively.
At the time interval $[t_0, t_2]$, $p_i$ never writes, and it
completes an SDC.  That is, $p_i$ reads the array twice, and in both
cases finds out that its preference (i.e., $\mypref_i$) appears in all
the $m$ registers.
There are three possible cases.
\begin{enumerate}
\item
\emph{Process $p_j$ does not write during the time interval $[t_0,t_1]$.}
Thus, since in the time interval $[t_0,t_1]$, $p_i$ has found
that the value of each one of the $m$ registers equals $v$,
it must be that at time $t_0$, the value of each one of the $m$
registers equals $v$.
After time $t_0$, and before executing line 4, process $p_j$ might write
at most once into one of the $m$ registers possibly overwriting the
$v$ value.  Thus, when executing line 5, $p_j$ will find that $v$
appears in at least $m-1$ of the entries of $\myview_j[1..m]$.  Since
$m\geq 3$, this means that $p_j$ will find that $v$ appears in more than
half of the entries of $\myview_j[1..m]$.  Thus, $p_j$ will set its
preference to $v$ (line 6).
From that point on, since $p_i$ does not write anymore,
the only possible decision value for $p_j$ is $v$.
\item
\emph{Process $p_j$ has written a value $u\neq v$ during the time
  interval $[t_0,t_1]$.}
Since in the time interval $[t_1,t_2]$, $p_i$
has found that the value of each one of the $m$ registers equals $v$,
it must be that after writing the value $u$, process $p_j$ has later
written the value $v$ (overwriting $u$). Thus, between these two
writings (of $u$ and later $v$), $p_j$ must have changed its preference
to $v$.  From that point on, since $p_i$ does not write anymore, the
only possible decision value for $p_j$ is $v$.
\item
  \emph{Process $p_j$ has written only the value $v$ during the time
    interval $[t_0,t_1]$.}  Let $t_0^\prime$ be the time after the
  last write of $p_j$ in $[t_0,t_1]$.  Since $p_i$ never writes after
  $t_0$, at time $t_0^\prime$ the values of all the registers written
  by $p_j$ are $v$. Also, the values of all the other registers must be
  $v$, since this is their value when $p_j$ reads them during
  $[t_0,t_1]$. Thus, $p_j$ will never change its preference after
  $t_0^\prime$, and the only possible decision value for $p_j$ is $v$.
\end{enumerate}
We have shown that both $p_i$ and $p_j$ decide on the same value $v$ in
$\rho$. Thus, the $n$ processes together decide on at most $n-1$
different values in $\rho$.

Next, we show that each process eventually decides (and terminates)
under obstruction-freedom (that is, if it runs alone for a
sufficiently long time).
When a process, say process $p_i$, runs alone
from some point on in a computation, $p_i$ will read the shared array
(line 4) and set its preference to some value $v$.  From that point
on, in each iteration of the repeat loop, $p_i$ will set one more entry
of the shared array to $v$. Thus, after at most $m$ iterations the
values of all the $m$ entries will equal $v$, and $p_i$ will be able to
exit the repeat loops, decide $v$ and terminate. \qed

\begin{lemma}[Validity]
  \label{lemma-sa-2}
The decision value is the input of a participating process.
\end{lemma}
\proof
At each point, the current preference of a process is either
its initial input or a value (different from $\bot$), it has read from
a register.  Since a process may only write its preference into a
register, the result follows. \qed

\begin{theorem}
  Algorithm 3 solves set agreement in a fully anonymous
  system made up of $n\geq 2$ processes and $m\geq 3$ anonymous RW registers.
\end{theorem}
\proof
The proof that the algorithm satisfies the Validity, Agreement,
and Obstruction-freedom properties (which define set agreement) follows
directly from Lemma~\ref{lemma-sa-1} and  Lemma~\ref{lemma-sa-2}. \qed

\section{Fully Anonymous 2-Process Obstruction-free Consensus using RW Registers}
\label{sec:ob-n=2-consensus}
\paragraph{A simple  instantiation of  Algorithm 3}
As the reader can easily check, instantiating Algorithm 3 with $n=2$
provides us with 2-process obstruction-free consensus built using
$m\geq 3$ RW registers.

\noindent
\textbf{Remark.}
Let us consider Algorithm 3, which assumes $n\geq 2$, in which
the requirement $m\geq 3$ is strengthened to $m\geq 2n-1$.
It is tempting to think that the resulting algorithm solves obstruction-free
consensus for $n$ processes, however, this is incorrect as the resulting 
algorithm does not even solve obstruction-free consensus for three processes
using five registers. Finding a counterexample
is left as an exercise for the reader.
%

Finally, it was recently proved in~\cite{T19} that there is no
obstruction-free consensus algorithm for two non-anonymous processes using only anonymous bits.
Thus, as was shown in~\cite{T19},
anonymous bits are strictly weaker than anonymous (and hence also
non-anonymous) multi-valued registers.

\section{Conclusions}
\label{sec:conclusion}
This article has several contributions. The first is the introduction
of the notion of {\it fully anonymous} shared memory
systems, namely, systems where the processes are anonymous and there
is no global agreement on the names of the shared registers (any
register can have different names for distinct processes). The article
has then addressed the design of a mutual exclusion
algorithm and agreement algorithms (consensus and set agreement)
in specific contexts where the anonymous registers are
read/write (RW) registers or more powerful  read/modify/write (RMW)
registers.
On the mutual exclusion side, the paper has shown that, for fully
anonymous mutual exclusion based on RMW registers, the condition on
the number $m$ of registers, namely $m\in M(n)= \{ m \mbox{ such that
} \forall \ell: 1< \ell \leq n: \gcd(\ell,m)=1\}$, is both necessary
and sufficient, extending thereby a result of~\cite{ARW18}
(which was for non-anonymous processes and anonymous registers).

Last but not least, let us notice that, despite the strong adversary context
(full anonymity, and failures in the case of agreement algorithms),
the proposed algorithms are relatively simple to
understand\footnote{Let us remind that simplicity is a first class
  property~\cite{AZ10,D80}. A stated by J. Perlis (the recipient of
  the first Turing Award) ``Simplicity does not precede complexity,
  but follows it''.}. However, some of their proofs are subtle.


\section*{Acknowledgments}
M. Raynal was partially supported by
the French ANR project DESCARTES  (16-CE40-0023-03) devoted to layered
and modular structures in distributed computing.
%

\end{document}